# Insulator to Semiconductor Transition in Graphene Quantum Dots

Himadri Sekhar Tripathi[1,a], Rajesh Mukherjee[2], Moumin Rudra[1], Ranjan Sutradhar[1], R. A. Kumar[1], T. P. Sinha[1]

[1]*Department of Physics, Bose Institute, 93/1, APC Road, Kolkata, India – 700009*
[2] *Department of Physics, Ramananda College, Bishnupur, Bankura, India –722122*

[a]Corresponding author: htripathi062@gmail.com

**Abstract.** Zero dimensional graphene quantum dots (GQDs) exhibit interesting physical and chemical properties due to the edge effect and quantum confinement. As the number of carbon atoms in edge is more than on basal plane, GQDs are more reactive. Room temperature XRD pattern confirms the formation of the GQDs. UV-Visible spectra confirm that GQDs show optical absorption in the visible region. The emission peaks in the photoluminescence spectra are red shifted with the increase of excitation wavelength. Dynamic light scattering (DLS) analysis shows that the average size of the particles is found to be ~65 nm. The frequency dependent electrical transport properties of the GQDs are investigated in a temperature range from 300 to 500 K. Most interestingly, for the first time, the insulator to semiconductor transition of GQD is observed near 400K. The transition mechanism of GQD is discussed with detailed dielectric analysis. The effects of intercalated water on temperature dependent conductivity are clearly discussed. The dielectric relaxation mechanism is explained in the framework of permittivity, conductivity and impedance formalisms. The frequency dependent ac conductivity spectra follows the Jonscher's universal power law. Cole-Cole model is used to investigate the dielectric relaxation mechanism in the sample.

## INTRODUCTION

Graphene quantum dots (GQDs), a new kind of quantum dots, have attracted extraordinary attention in current research on nanoscience and nanotechnology due to their outstanding properties, such as high surface area, good solubility, very poor cytotoxicity, stable fluorescence and adjustable bandgap [1]. These properties recommend GQDs for applications in bio imaging [2–7], bio- and metal sensing [8–12], photovoltaic [13–15], and photo catalysts [16-17] etc. On the other hand, GQDs having size in nanometer scale with $sp^2$-$sp^2$ carbon bonds show characteristic properties such as size dependent quantum dots (QDs) [18-20]. Most applications of GQDs are focused on the photoluminescence (PL)-related fields since it shows different PL spectra for different size of the particles. Again, due to their small size GQDs are well dispersible in different organic solvents. Modifying the different organic solvents and synthesis procedure, different kind of PL color can be observed, which indicates the changes in the band gap. Recently, additional excellent properties of GQDs such as high transparency and high surface area have been proposed for energy storage applications. Because of its large surface area, electrodes using GQDs are applied to capacitors [21] and batteries [22-23], and the conductivity of GQDs is higher than that of graphene oxide (GO) [24]. Despite of these efforts, there are some important challenges for studies on the electrical properties of GQDs. For example, a detailed study of the effect of temperature on conductivity and dielectric properties is still absent. Moreover, the roles of intercalated water in GQDs essentially need for exact clarifications. Most interestingly, we have studied the dielectric spectroscopy in wide temperature interval to investigate temperature dependent electrical properties of GQDs. Again, the insulator to semiconductor transition in GQDs has been reported for the first time and the transition mechanism has been clearly demonstrated by analyzing Cole-Cole model.

## EXPERIMENTAL

### Synthesis

GQDs were prepared by simple pyrolysis of citric acid [24] with some modifications. Five grams of citric acid were heated and melted at 338 K - 343 K temperature. This melted citric acid was converted into dark orange color within 25 – 30 min. Then this dark citric acid solution was kept at room temperature. 2 M solution of NaOH was added drop wise in the melted dense solution of citric acid at room temperature. PH of the solution was tested for several times. Finally, the GQDs are prepared at PH level 11. Drying of this GQDs solution was a very important step. The temperature of the solution was increased slowly and was dried at temperature ~323K. GQDs were collected in powder form in a container.

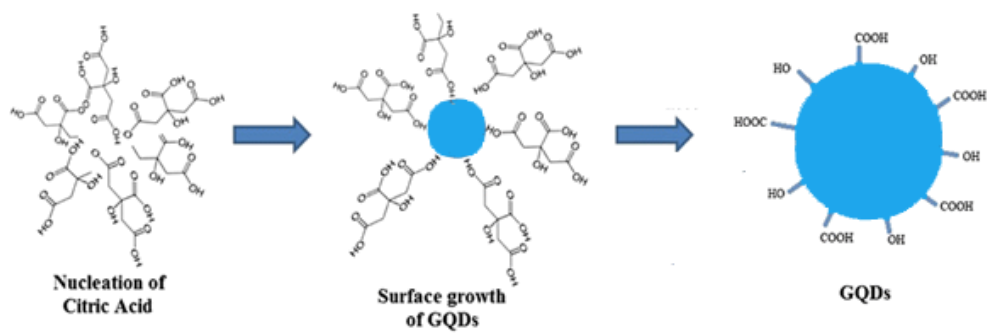

**FIGURE 1**: Formation of Graphene Quantum Dots (GQDs)

## Measurement

Room temperature X-ray diffraction (XRD) pattern of the sample is carried out from Bragg angle $2\theta = 10^0$ to $40^0$ with the Rigaku Miniflex II diffractometer having Cu-K$_\alpha$ radiation ($\lambda$ = 0.1542 nm). The Dynamic Light Scattering (DLS) spectrum was recorded by Malvern Zetasizer Nano ZS. The UV–Visible spectrum of the materials was collected at room temperature using UV–Vis spectrophotometer (Shimadzu UV 2401Pc). The photoluminescence (PL) spectra of GQDs were taken by JASCO FP-8500. Impedance spectroscopy of GQDs was performed in the temperature range 300K to 500K and the frequency range from 45 Hz to 5 MHz of the oscillating voltage of 1.0 V utilizing an LCR meter (HIOKI – 3532, Japan). Eurotherm 818p programmable temperature controller was used to control the furnace temperature with an accuracy of ±1 K. Before the experiment, the flat surfaces on both sides of the pellet were cleaned properly and contacts were made from the thin silver paste.

## RESULTS AND DISCUSSIONS

### Structural Analysis

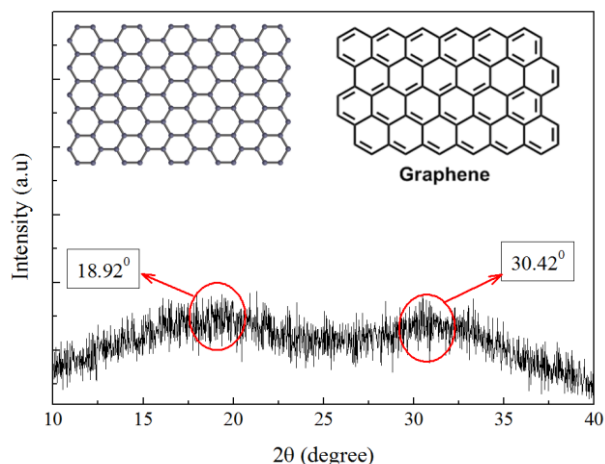

**FIGURE 2:** X-ray diffraction pattern of GQDs

**FIGURE 2** shows the XRD pattern of GQD which exhibits a diffraction peak at $2\theta = 18.42^0$ which corresponds to the inter layer spacing 0.468 nm. Again, another diffraction peak at $2\theta = 30.42^0$, corresponding the *d* spacing 0.293 nm which attributes the (002) plane [1]. The *d* value of GQDs is broader than graphite which suggests more oxygen containing functional group attached to GQDs [2]. Furthermore, the *d* value of GQDs is less than graphene oxide which could be explained that the functional groups are attached to the edges of GQDs rather than the basal plane [2]. However, the absence of a sharp peak in the XRD pattern of GQDs is due to the disorder structure of GQDs. This conclusion confirms about the presence of oxygenated functional groups which can play the considerable effect in the microstructure of GQDs [7]. **FIGURE 3** shows the dynamic light scattering (DLS) spectra of as synthesized GQDs. The spectra confirm that the average size of GQDs is approximately 66.82 nm.

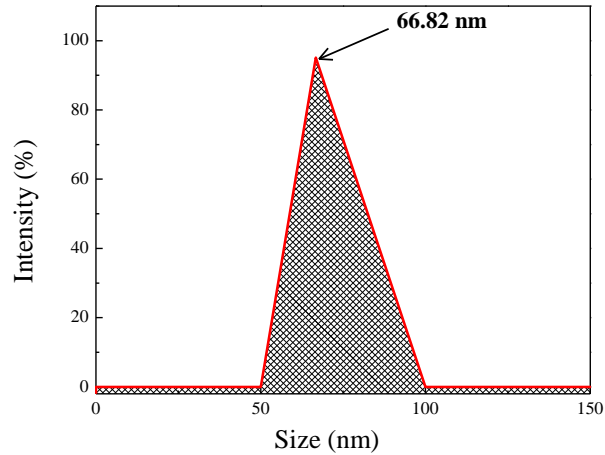

**FIGURE 3:** Dynamic light scattering spectra of GQDs

## Optical Analysis

**FIGURE 4 (a)** shows the Photoluminescence (PL) spectra of GQDs with excitation wavelength ranges from 440 nm to 480 nm. GQDs exhibit clear excitation dependent PL behavior. It is seen that PL peak shifts towards higher excitation wavelength with decreasing intensity. At 440 nm excitation PL spectrum shows a strong peak at 513 nm.

UV-visible absorption spectrum was used to determine the band gap of the GQDs and the obtained results are shown in **FIGURE 4 (b)**. The band gap is determined using the Tauc relation [26], $(\alpha h\nu)^{1/n} = C(h\nu - E_g)$, where $E_g$ = Band gap, $\alpha$ is the optical absorption coefficient, $h$ is the Planck constant, $\nu$ is the frequency of light photons, $C$ = constant, $n$ depends on the type of transition ($n = 1/2$ for direct allowed transition, $n = 2$ for indirect allowed transition, $n = 3/2$ for direct forbidden transition, $n = 3$ for indirect forbidden transition). The value of the optical band gap (already indexed) for the direct optical transition is found to be 2.82 eV for GQD.

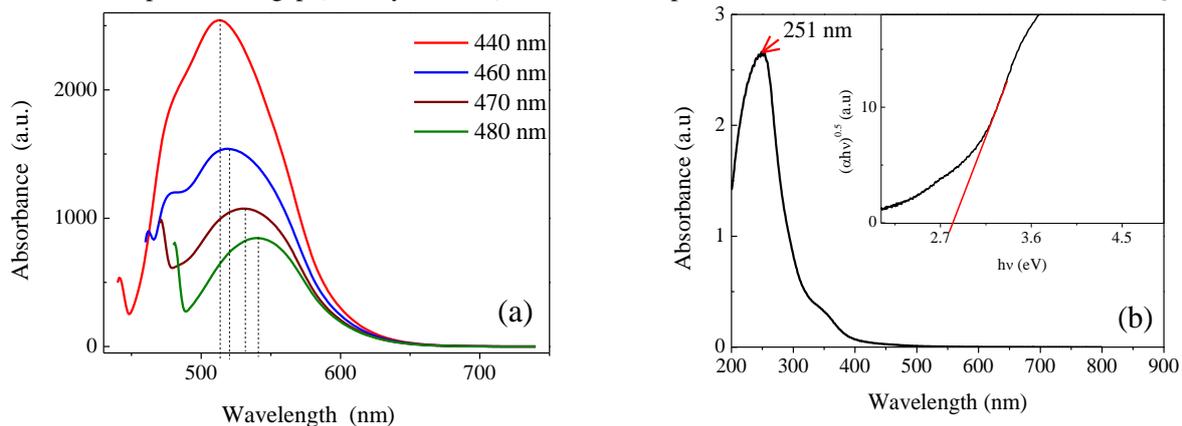

**FIGURE 4:** a) Photoluminescence (PL) of GQDs at different excitations and b) UV-visible spectra of GQDs.

## ELECTRICAL IMPEDANCE ANALYSIS

### Conductivity analysis

Ac conductivity spectra are an important tool to study the electrical conduction mechanism in the synthesized GQDs. **FIGURE 5 (a)** shows the frequency dependent ac conductivity spectra of GQDs in various temperature regions. In the conductivity spectra, there are two distinct regions, region I is the frequency independent part (plateau region), i.e. dc conductivity ($\sigma_{dc}$) and region II is dispersion region, where the ac conductivity increases with the increasing frequency. The variation of conductivity can be explained by the well-known Johnscher's power law [27].

$$\sigma_{ac} = \sigma_{dc} + A\omega^n \tag{1}$$

Here, $\omega$ (=$2\pi f$) is the angular frequency, the coefficient $A$ and the exponent $n$ ($0 < n < 1$) are the temperature and material dependent constants. The values of the fitted parameters are listed in **TABLE1**.

**FIGURE 5 (b)** shows the variation of logarithmic dc conductivity with increasing temperature. Above 380 K the GQDs shows insulating behavior and below it semiconductor behavior. At room temperature, the water molecules intercalated in the microstructure of GQDs are immobile. As the temperature increases, water molecules absorb heat and slowly become mobilized. Then, the water molecule starts to disappear from the microstructure of GQDs and then the conductivity increases. Again in the semiconductor region, there appears slight fluctuations of conductivity with respect to temperature due to some functional group at the edge in GQDs.

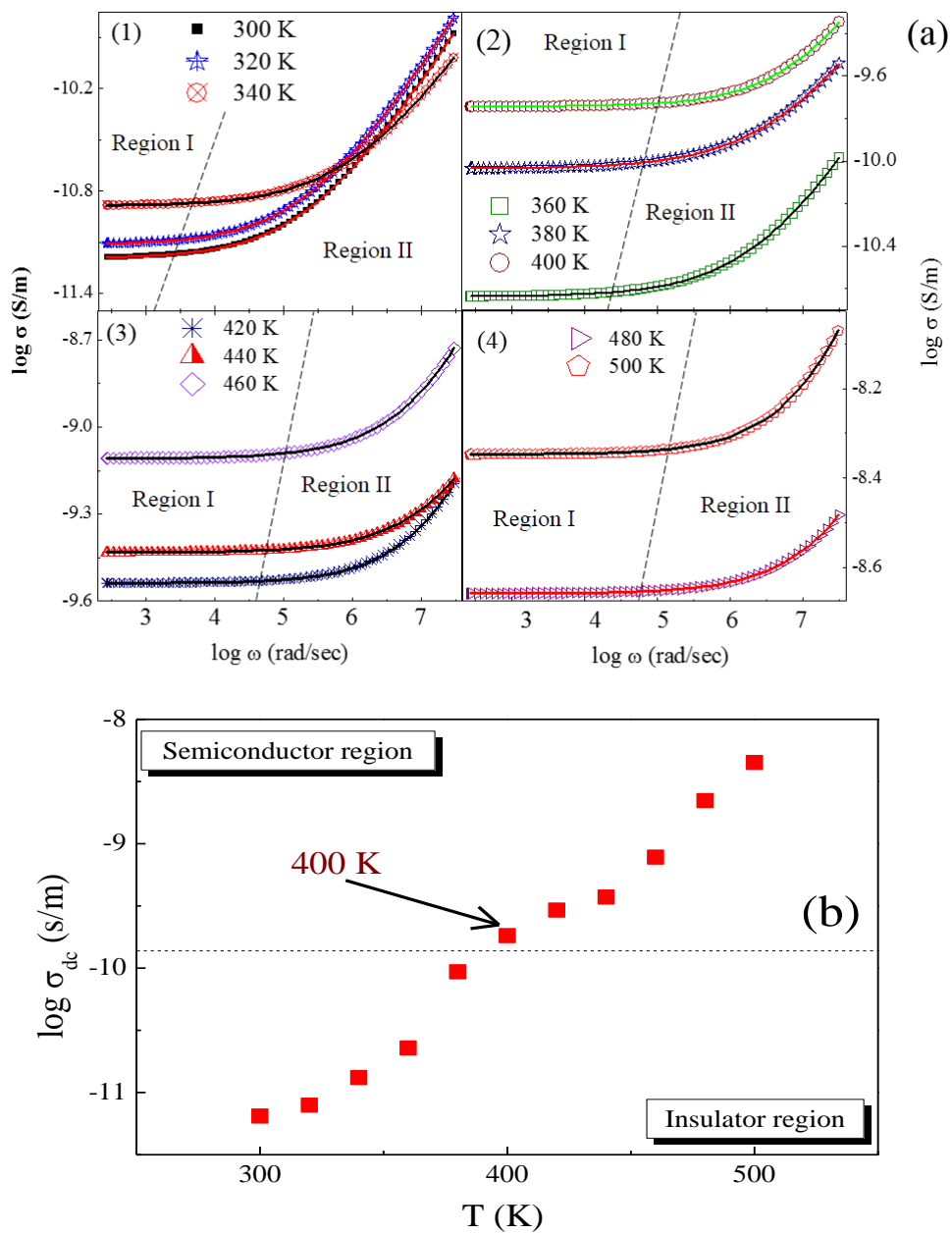

**FIGURE 5:** a) The frequency dependence of ac conductivity (dotted line) at various temperatures with fitted data (solid line) and b) the variation of dc conductivity with temperature.

TABLE 1. Fitting parameter of ac conductivity

| Temperature, T (K) | $\sigma_{ac} = \sigma_0 + A\omega^n$ | | |
|---|---|---|---|
| | $\sigma_0$ (X$10^{-11}$) (S/m) | A (X$10^{-15}$) | n |
| 300 | 6 | 2.9 | 0.62 |
| 320 | 0.77 | 3.6 | 0.62 |
| 340 | 1.3 | 3.8 | 0.58 |
| 360 | 2.3 | 3.1 | 0.59 |
| 380 | 9.2 | 15 | 0.55 |
| 400 | 18 | 4.4 | 0.64 |
| 420 | 29 | 4.0 | 0.66 |
| 440 | 37 | 9.5 | 0.60 |
| 460 | 78 | 25 | 0.62 |
| 480 | 220 | 25 | 0.62 |
| 500 | 450 | 55 | 0.65 |

## Dielectric Relaxation

**FIGURE 6 (a)** shows the frequency dependent dielectric constant ($\varepsilon'$) in the temperature range from 300 K to 500 K. At low frequency permanent dipoles align along the field direction and dielectric constant is high. After that it decreases with increasing frequency and becomes constant at high frequency because electric dipoles can no longer follow the applied electric field at high frequency region. Again, the dielectric constant increases with temperature as the conductivity increases due to removal of intercalated water molecules in the GQDs.

According to Cole-Cole model complex dielectric constant ($\varepsilon^*$) can be written as

$$\varepsilon^* = (\varepsilon' - \varepsilon'') = \varepsilon_\infty + (\varepsilon_s - \varepsilon_\infty)/(1 + (j\omega\tau)^{1-\alpha}) \qquad (2)$$

Where $\varepsilon'$ is the real part of $\varepsilon^*$, $\varepsilon''$ is the imaginary part of $\varepsilon^*$, $\varepsilon_s$ is the static dielectric constant, $\varepsilon_\infty$ is the value of dielectric constant at infinity. $(\varepsilon_s - \varepsilon_\infty)$ is called dielectric strength, $\tau$ is the mean relaxation time and α is the degree of the distribution time which varies from 0 to 1 (α = 0 corresponds to standard Debye relaxation). From above equation we evaluate the value of ε' as

$$\varepsilon' = \varepsilon_\infty + \frac{(\varepsilon_s - \varepsilon_\infty)\{1 + (\omega\tau)^{1-\alpha}\sin(\frac{\alpha\pi}{2})\}}{1 + 2(\omega\tau)^{1-\alpha}\sin(\frac{\alpha\pi}{2}) + (\omega\tau)^{2(1-\alpha)}} \qquad (3)$$

A moderately good fitting of the experimental data using equation (3) for different temperatures (300 K to 500 K) are shown in the Fig 6 by solid lines and the values of fitting parameters for different temperature are listed in **TABLE2**
The value of α is zero for an ideal Debye relaxation. Here α > 0, which leads to a broader peak than a Debye peak.

TABLE 2. Fitting parameter of dielectric constant

| Temperature, T (K) | Fitting parameter of $\varepsilon'$ | | | |
|---|---|---|---|---|
| | $\varepsilon'_\infty$ | $\varepsilon'_S - \varepsilon'_\infty$ (X$10^9$) | $\tau$ | $\alpha$ |
| 300 | 700.1202 | 4.69 | 0.1195 | 0.2 |
| 320 | 700.1202 | 12.69 | 0.1195 | 0.2 |
| 340 | 700.1202 | 44.69 | 0.1195 | 0.2 |
| 360 | 700.1202 | 156.69 | 0.1195 | 0.2 |
| 380 | 700.1202 | 1000 | 0.1195 | 0.2 |
| 400 | 700.1202 | 2906.9 | 0.1195 | 0.2 |
| 420 | 700.1202 | 3050 | 0.1195 | 0.2 |
| 440 | 700.1202 | 1800 | 0.1195 | 0.2 |
| 460 | 700.1202 | 1400 | 0.1195 | 0.30 |
| 480 | 700.1202 | 2500 | 0.1195 | 0.32 |
| 500 | 700.1202 | 4500 | 0.1195 | 0.35 |

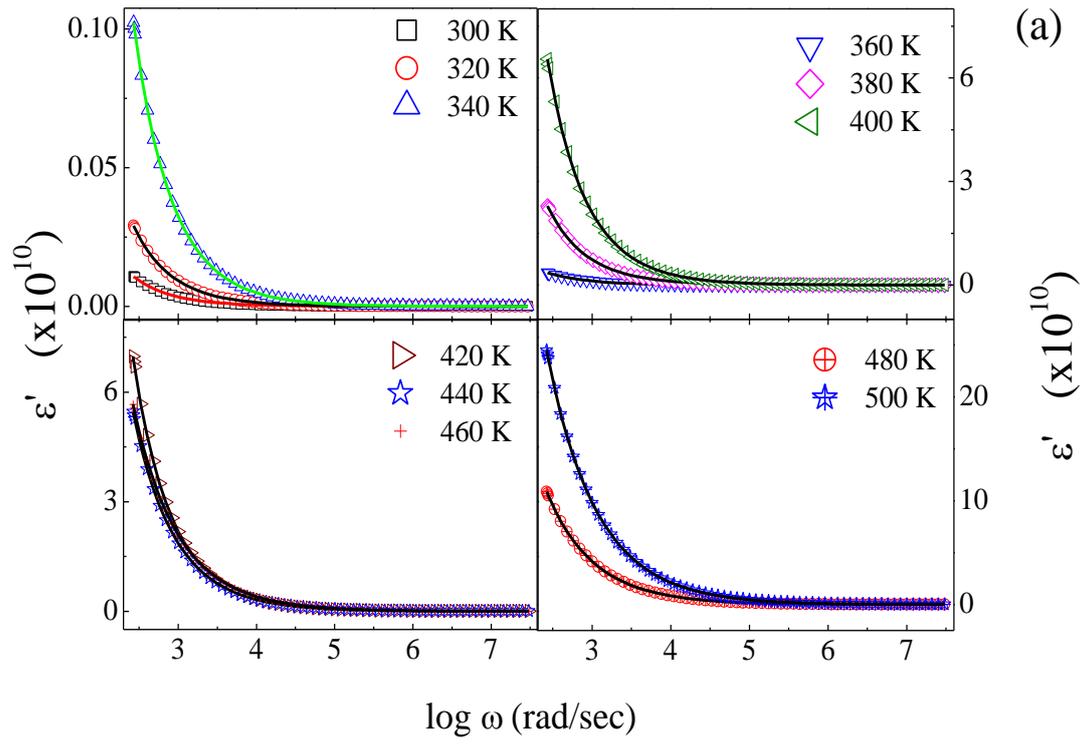

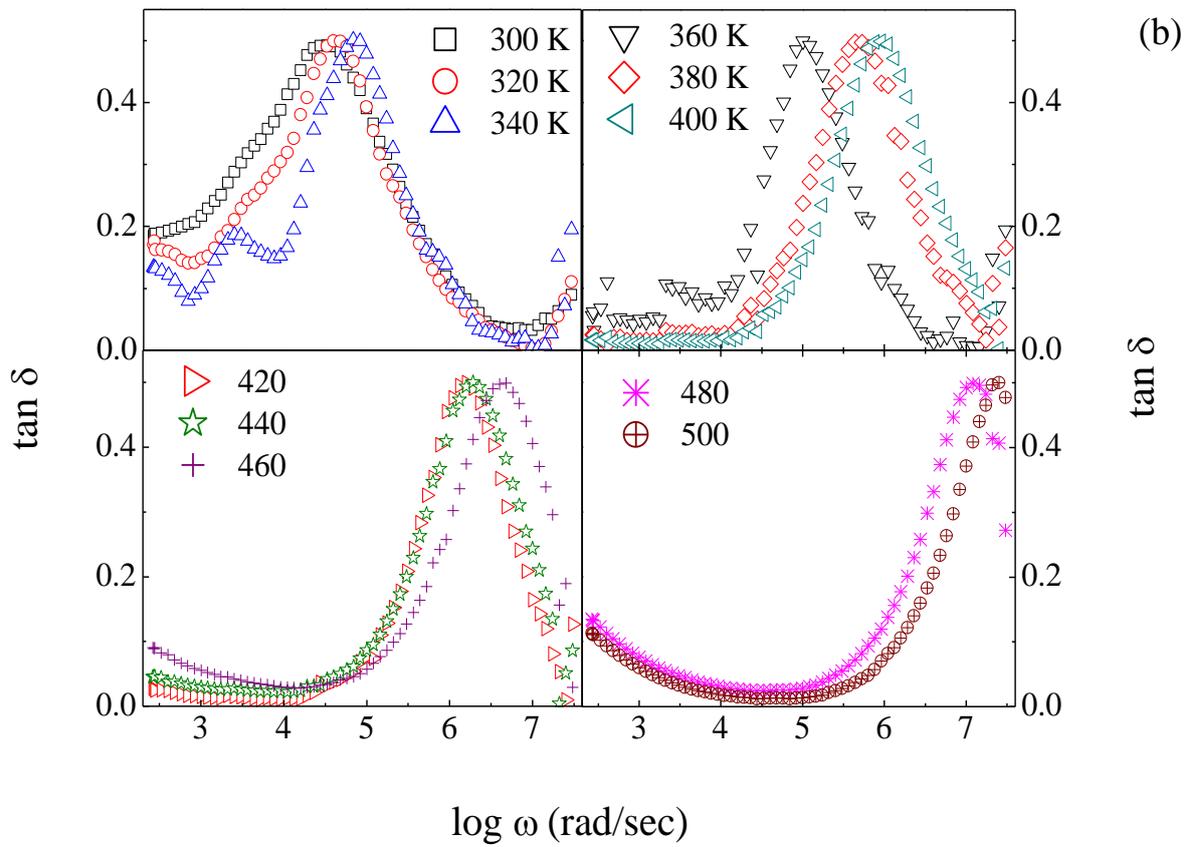

**FIGURE 6:** a) Real part of dielectric constant at different temperature, and b) the frequency dependence of tan$\delta$ at various temperatures.

The relaxation peaks develop in the loss tangent (tanδ) at the dispersion region of ε' is shown in the **FIGURE 6 (b)**. The peak position in tanδ shifts to the high frequency side with increasing temperature. At high temperature rate of polarization is high and relaxation occurs at high frequency.

## Impedance spectra

**FIGURE 7 (a)** shows complex impedance plane plots for GQDs at different temperatures from 300 K to 500 K. For every temperature, there exists only one semicircular arc which indicates one type of relaxation process to the charge transport of GQDs. Due to the non-ideal behavior of capacitance, the center of the semicircles is found depressed below the real axis. An RC equivalent circuit is replaced by a constant phase element (CPE) to accommodate the non-ideal behavior of the capacitance. The capacitance of CPE can be written as $C_{CPE} = Q^{1/n} R^{(1-n)/n}$, where $n$ estimates the non-ideal behavior. The value of $n$ is zero for the ideal resistance and unity for the ideal capacitance. Here $R$ and $Q$ are the resistance and constant phase element, respectively. The solid lines in Fig. 8 represent the fitting to the electrical equivalent circuit and the fitted parameter $R$, $Q$ and $n$ were obtained for different temperatures are listed in **TABLE 3**.

**TABLE 3.** Fitting parameter of Z´ and Z´´

| | Fitting parameters of Z´ and Z´´ | | |
|---|---|---|---|
| **Temperature, T (K)** | $R_e$ (×$10^5$) (Ω) | $Q_e$ (×$10^{-10}$) | $n_e$ |
| 300 | 11 | 1.2799999 | 0.85 |
| 320 | 8 | 1.2799999 | 0.85 |
| 340 | 6 | 1.2799999 | 0.82 |
| 360 | 3 | 1.2799999 | 0.81 |
| 380 | 50 | 1.2799999 | 0.81 |
| 400 | 40 | 1.2799999 | 0.81 |
| 420 | 20 | 1.2799999 | 0.81 |
| 440 | 20 | 1.2799999 | 0.81 |
| 460 | 1500 | 1.2799999 | 0.81 |
| 480 | 1000 | 1.2799999 | 0.81 |
| 500 | 700 | 1.2799999 | 0.81 |

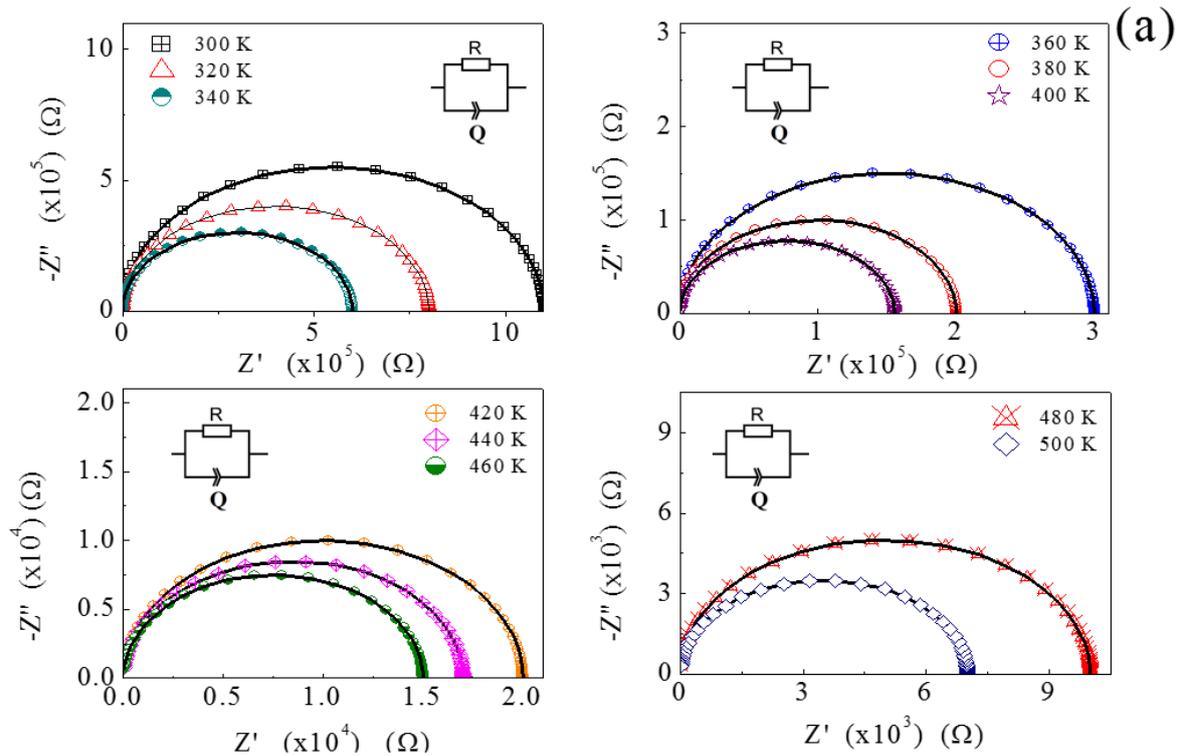

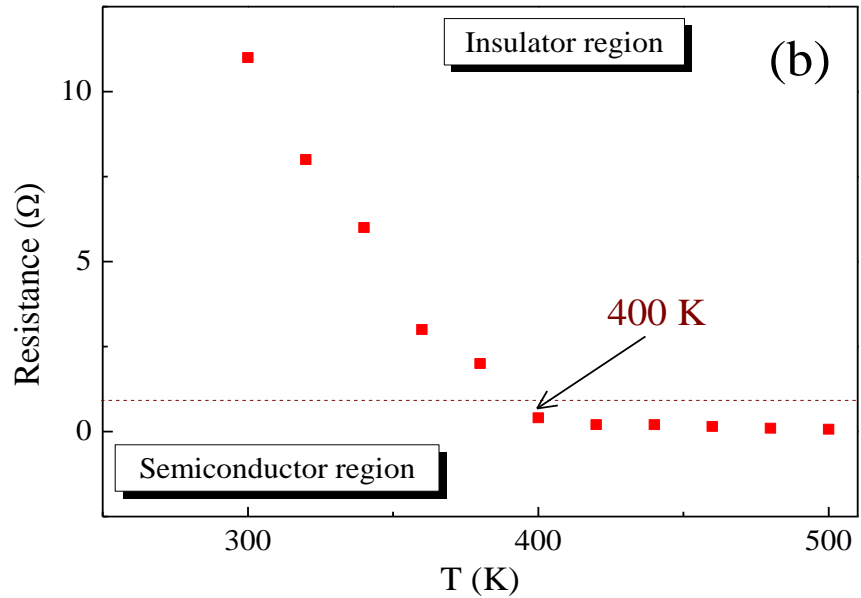

**FIGURE 7**: a) Nyquist plot at different temperature, insets show the equivalent circuit used and b) Resistance of CR circuit versus temperature.

The variation of resistance in increasing of temperature is shown in the **FIGURE 7 (b)**. The increase in diameter of the semicircular arcs with the decrease of temperature points towards the thermally activated conduction mechanism in GQDs at the electrode. The value of resistance decreases with increasing temperature, suggesting the thermal activation of the localized charges [34].

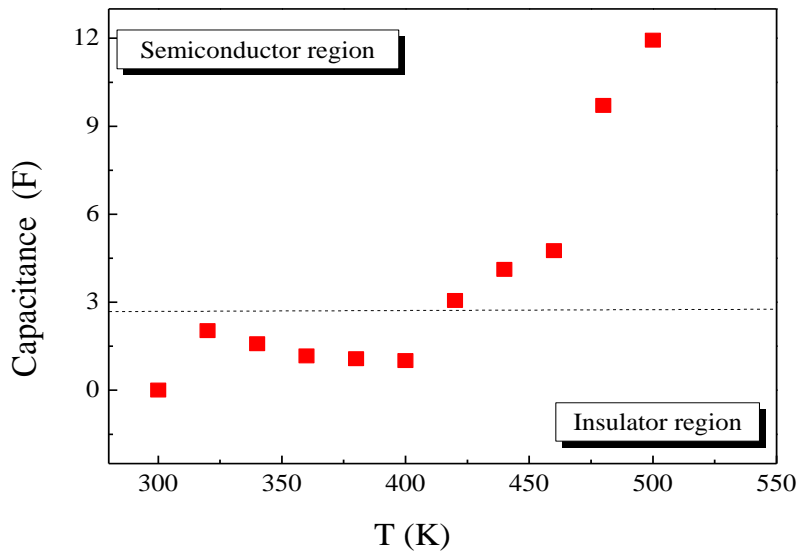

**FIGURE 8**: The variation of capacitance with temperature

**FIGURE 8** shows the capacitance variation with increasing temperature. Since at room temperature there are some immobile water molecules in the GQDs, the capacitance of GQDs is almost equal from 300 K to 400 K. The capacitance increases rapidly above 400 K due to removal of water molecules from the microstructure of GQDs.

## CONCLUSION

For the first time, an in situ measurement was designed to investigate the electrical property evolution of GQDs at a temperature range bet 300 K to 560 K. A clear insulator to semiconductor transition was demonstrated. The insulator– semiconductor transition was observed at about 400 K and the detailed transition mechanism was confirmed by dielectric properties and microstructure analyses. Our work on GQDs provides a comprehensive understanding of the electrical properties, structures and chemical bonds of GQDs. Considering the unique properties of GQDs; this progress may be very useful for quality improvement in the fabrication of flexible electronic devices etc.

## ACKNOWLEDGEMENT

Himadri Sekhar Tripathi (Inspire ID – IF170057), Ranjan Sutradhar (IF160017) acknowledge the financial support provided by the Department of Science and Technology (DST), New Delhi, 110016. Moumin Rudra (ID 522407) thanks the University Grants Commission (UGC), New Delhi for providing financial support with award letter no. 2121551156 dated 21.06.2016. R. A. Kumar acknowledges the financial support provided by the Council of Scientific and Industrial Research (CSIR), New Delhi.

# REFERENCES

[1]. L. L. Li, G. H. Wu, G. H. Yang, J. Peng, J. W. Zhao and J. J. Zhu, Nanoscale **5**, 4015–4039 (2013).
[2]. S. J. Zhu, J. H. Zhang, C. Y. Qiao, S. J. Tang, Y. F. Li, W. J. Yuan, B. Li, L. Tian, F. Liu and R. Hu, Chem. Commun **47**, 6858–6860 (2011).
[3]. S. J. Zhu, J. H. Zhang, S. J. Tang, C. Y. Qiao, L. Wang, H. Y. Wang, X. Liu, B. Li, Y. F. Li and W. L. Yu, Adv. Funct. Mater **22**, 4732–4740 (2012).
[4]. J. Peng, W. Gao, B. K. Gupta, Z. Liu, R. Romero-Aburto, L. H. Ge, L. Song, L. B. Alemany, X. B. Zhan, G. H. Gao, S. A. Vithayathil, B. A. Kaipparettu, A. A. Marti, T. Hayashi, J. J. Zhu and P. M. Ajayan, Nano Lett. **12**, 844–849 (2012).
[5] L. Wang, Y. L. Wang, T. Xu, H. B. Liao, C. J. Yao, Y. Liu, Z. Li, Z. W. Chen, D. Y. Pan, L. T. Sun and M. H. Wu, Nat. Commun **5**, 5357 (2014).
[6] X. Wu, F. Tian, W. X. Wang, J. Chen, M. Wu and J. X. Zhao, J. Mater. Chem. C **1**, 4676–4684 (2013).
[7] D. Y. Pan, L. Guo, J. C. Zhang, C. Xi, Q. Xue, H. Huang, J. H. Li, Z. W. Zhang, W. J. Yu and Z. W. Chen, J. Mater. Chem. **22**, 3314–3318 (2012).
[8] S. Umrao, M. H. Jang, J. H. Oh, G. Kim, S. Sahoo, Y. H. Cho, A. Srivastva and I. K. Oh, Carbon **81**, 514–524 (2015).
[9] Y. Y. Liu and D. Young´aKim, Chem. Commun. **51**, 4176–4179 (2015).
[10] J. J. Liu, Z. T. Chen, D. S. Tang, Y. B. Wang, L. T. Kang and J. N. Yao, Sens. Actuators B **212**, 214–219 (2015).
[11] A. X. Zheng, Z. X. Cong, J. R. Wang, J. Li, H. H. Yang and G. N. Chen, Biosens. Bioelectron. **49**, 519–524 (2013).
[12] Y. H. Zhu, G. F. Wang, H. Jiang, L. Chen and X. J. Zhang, Chem. Commun. **51**, 948–951 (2015).
[13] X. Yan, X. Cui, B. S. Li and L. S. Li, Nano Lett. **10**, 1869–1873 (2010).
[14] Y. Li, Y. Hu, Y. Zhao, G. Q. Shi, L. Deng, Y. B. Hou and L. T. Qu, Adv. Mater **23**, 776–780 (2011).
[15] Z. C. Huang, Y. T. Shen, Y. Li, W. J. Zheng, Y. J. Xue, C. Q. Qin, B. Zhang, J. J. Hao and W. Feng, Nanoscale **6**, 13043–13052 (2014).
[16] H. T. Li, X. D. He, Z. H. Kang, H. Huang, Y. Liu, J. L. Liu, S. Y. Lian, C. H. A. Tsang, X. B. Yang and S. T. Lee, Angew. Chem., Int. Ed. **49**, 4430–4434 (2010).
[17] S. J. Zhuo, M. W. Shao and S. T. Lee, ACS Nano **6**, 1059–1064 (2012).
[18] Z. Zhang, J. Zhang, N. Chen, L. Qu, Energy Environ. Sci. **5**, 8869-8890 (2012).
[19] R. Sekiya, Y. Uemura, H. Murakami, T. Haino, Angew. Chem. Int. Ed. **53** 5619-5623 (2014).
[20] J. Shen, Y. Zhu, X. Yang, C. Li, Chem. Commun. **48**, 3686-3699 (2012).
[21] W. Liu, Y. Feng, X. Yan, J. Chen, Q. Xue, Adv. Funct. Mater **23**, 4111-4122 (2013).
[22] D. Chao, C. Zhu, X. Xia, J. Liu, X. Zhang, J. Wang, et al., Nano Lett. **15**, 565-573 (2014).
[23] D. Pan, C. Xi, Z. Li, L. Wang, Z. Chen, B. Lu, et al., J. Mater. Chem. A **1**, 3551-3555 (2013).
[24] X. Yan, X. Cui, L. Li, J. Am. Chem. Soc. **132**, 5944-5945 (2010).
[25] J. P. Naik, P. Sutradhar, M. Saha, J. Nanostruct Chem. **7**, 85–89 (2017).
[26] J. Tauc, R. Grigorovici, and A. Vancu, Phys. Status Solidi **15,** 627 (1966*)*.
[27] A. K. Jonscher, *Nature* **267**, 673–679 (1977).
[28] K. S. Cole, R. H. Cole, J. Chem. Phys. **10**, 98 (1942).